\documentclass[aps,pre,nofootinbib,showpacs,twocolumn]{revtex4-1}%
\usepackage{amssymb,latexsym,mathrsfs}
\usepackage{amsfonts}
\usepackage{mathrsfs}
\usepackage{amsmath}
\usepackage{graphicx}
\usepackage{color}
\usepackage{hyperref}

\newcommand{\beq}{\begin{equation}}
\newcommand{\eeq}{\end{equation}}
\newcommand{\beqn}{\begin{eqnarray}}
\newcommand{\eeqn}{\end{eqnarray}}
\newcommand{\bearr}{\begin{array}}
\newcommand{\enarr}{\end{array}}

\def\bea{\begin{eqnarray}}
\def\eea{\end{eqnarray}}
\def\ba{\begin{array}}
\def\ea{\end{array}}

\begin{document}
\title{Energy transport in the presence of long-range interactions}
\author{Debarshee Bagchi}
\email[E-mail address:]{debarshee@cbpf.br}
\affiliation{Instituto de F\'{\i}sica, Universidade Federal do Rio Grande do Sul, Porto Alegre RS, Brazil}
\date{\today}

\begin{abstract}
We study energy transport in the paradigmatic Hamiltonian mean-field (HMF) model and other related long-range interacting
models using molecular dynamics simulations. We show that energy diffusion in the HMF model is subdiffusive in nature, which
confirms a recently obtained intriguing result that, despite being globally interacting, this model is a thermal insulator
in the thermodynamic limit. Surprisingly, when additional nearest neighbor interactions are introduced to the HMF model, an
energy superdiffusion is observed. We show that these results can be consistently explained by studying energy localization
due to thermally generated intrinsic localized excitation modes (discrete breathers) in nonlinear discrete systems. Our
analysis for the HMF model can also be readily extended to more generic long-range interacting models where the interaction
strength decays algebraically with the (shortest) distance between two lattice sites. This reconciles many of the apparently
counter-intuitive results presented recently [C. Olivares and C. Anteneodo, Phys. Rev. E {\bf 94}, 042117 (2016); D. Bagchi,
Phys. Rev. E {\bf 95}, 032102 (2017)] concerning energy transport in two such long-range interacting models.
\end{abstract}


\maketitle


\section{Introduction}
\label{introduction}

Long-range (LR) interactions are ubiquitous at all length scales - from cosmology \cite{cosmo} to nanoscience \cite{nano} - and are 
being investigated extensively in recent times. These systems possess extremely rich dynamical and thermodynamical properties that
often deviate fantastically from short-range interacting systems. Very frequently these systems exhibit non-ergodicity, weak chaos, 
in-equivalence of ensembles, long-lived non-Gaussian quasistationary states, phase transitions in one dimension, non-concave entropy,
and negative specific heat (for reviews see Refs. \cite{LR_Levin,LR_Mukamel,LR_Ruffo,Ruffo2017Rev}), and as such the conventional
thermodynamical formalism of Boltzmann-Gibbs statistical mechanics becomes invalid. These unusual features make the study of LR 
interacting systems extremely interesting as well as challenging. 

Very recently two studies of energy transport in nonlinear one dimensional models with LR interactions have been performed. 
One of these is for the long-range Fermi-Pasta-Ulam (LR-FPU) model \cite{LRFPU} described by the Hamiltonian
\begin{equation}
\mathcal{H} = \sum_{i=1}^N \left[\frac{p_i^2}{2} + \frac 12 (x_{i+1} - x_i)^2 
+ \frac{1}{8 \tilde N} \sum_{j=1}^N \frac {(x_j - x_i)^4}{{|i-j|}^{\delta}}\right]
\label{H1}
\end{equation}
and the other is the long-range inertial XY (LR-XY) model \cite{LRXY} (we use the symbol $\delta$ here instead of $\alpha$
used in Ref. \cite{LRXY})
\begin{equation}
\mathcal{H} = \sum_{i=1}^N \left[\frac{p_i^2}{2} + \frac{1}{2\tilde N} \sum_{j=1}^N \frac {1 - \cos(\theta_j - \theta_i)}{{|i-j|}^{\delta}}\right],
\label{H2}
\end{equation}
where the symbols have their usual meaning. We have set all the coupling constants and mass of each particle to unity. The factor $ 1/\tilde N $
in the LR term of both the Hamiltonians ensures extensivity of the potential energy and depends on the parameter $\delta$, spatial dimension $d$
and system size $N$ \cite{LRFPU,LRXY}.  

The Hamiltonians of these two models resemble each other in the sense that the nonlinear interaction term in both has been modified in
a similar fashion to include LR interactions. Thus each particle interacts with all the other particles in the system and the strength
of interaction decays algebraically with the (shortest) distance between the two lattice sites, say $i$ and $j$, as $|i-j|^{-\delta}$.
The systems described by the Hamiltonians Eqs. (\ref{H1}) and (\ref{H2}) are considered to be long-ranged for $0 \le \delta < 1$ and
short-ranged if $\delta > 1$; for $\delta \to \infty$ we have the nearest-neighbor (NN) interacting models, namely, the Fermi-Pasta-Ulam (FPU)
and the inertial XY (coupled rotor) model whereas $\delta = 0$ corresponds to the mean-field scenario.

In energy transport studies of these two models it was observed that the LR-FPU model exhibits superdiffusive transport for all values
of the parameter $\delta$ \cite{LRFPU} whereas the LR-XY model exhibits two distinct phases - an insulator phase for $\delta < 1$ and a 
conducting phase $\delta > 1$ in which Fourier law (implying normal energy diffusion) is obeyed \cite{LRXY}. Moreover, in the LR-FPU, the
conductivity $\kappa$ has an intriguing non-monotonic dependence on $\delta$ with a maximum conductivity at $\delta = 2$ whereas the 
conductivity for the LR-XY model increases monotonically as one increases $\delta$ from zero.

These results are quite interesting and raise a few questions that need to be addressed. First, why are the transport properties of
these two LR systems so strikingly different from each other? In the thermodynamic limit, the LR-FPU is always a thermal conductor
(anomalous) with diverging heat conductivity whereas the LR-XY is a thermal insulator for small $\delta < 1$ and becomes a conductor (normal)
obeying Fourier's law for large $\delta > 1$.

Second, what is the microscopic mechanism responsible for the existence of an insulator phase in a classical Hamiltonian model where the
particles interact via LR interactions? On the contrary, one would expect that, in the presence of global interactions, energy would propagate
from one part of the lattice to the other extremely fast. In fact, very recently, an efficient model of a thermal diode was proposed 
\cite{LRDiode} that relied on the idea that LR interactions create additional channels favoring energy flow which may prevent the usual
decay of rectification in the thermodynamic limit that typically happens in short-range systems. 
This idea of increasing the energy flow by LR interactions was previously proposed and analytically investigated, but with an
infinitesimal temperature gradient across the system \cite{RectLR}. In another recent work \cite{RectNNN} it has been shown that the
addition of next-nearest-neighbor interactions is already enough to increase the energy transport and thermal rectification in a
mass-graded chain of anharmonic oscillators.
In striking contrast, from the study of the LR-XY model it was concluded that the thermal conduction is ``spoiled" in the presence of
LR interactions \cite{LRXY}.
This is the central puzzle that we will investigate here - why the introduction of LR interactions in the FPU model enhances its conductivity
significantly whereas doing the same in the XY model, quite surprisingly, turns it into a thermal insulator.

Third, the presence of an insulator phase in the thermodynamic limit demands that the underlying diffusion process should be subdiffusive
in nature. This in itself is remarkable since, to the best of our knowledge, apart from one example of a billiard channel model \cite{billiard},
there are no known classical Hamiltonian models that exhibit energy subdiffusion, in the absence of disorder. The well-known one-dimensional
Hamiltonian models that have been studied so far exhibit either ballistic transport (seen in integrable models such as the ordered Harmonic
lattice), superdiffusive transport (seen in linear momentum conserving systems such as the FPU model), or normal transport (seen in the coupled
rotor model and systems with on-site potentials such as the $\phi^4$ model), but an insulator in the thermodynamic limit is not usually
encountered \cite{Subdiff1, Subdiff2}. In fact, as a consequence of the Kac lemma, subdiffusion is argued to be forbidden in conventional
Hamiltonian systems due to finiteness of the Poincar\'{e} recurrence time \cite{Kaclemmabook}, although some special cases of subdiffusion
have been observed \cite{ZasRev}.

In this paper, we attempt to obtain a qualitative and, wherever possible, quantitative understanding of the issues raised above. 
The remainder of the paper is organized as follows. In Sec. \ref{sec:model} we describe the paradigmatic Hamiltonian mean-field 
(HMF) model \cite{LR_Levin,LR_Mukamel,LR_Ruffo,Ruffo2017Rev} which is the mean-field ($\delta=0$) limit of the LR-XY system
described by Eq. (\ref{H2}). We study numerically the HMF model (and some other closely related models) to first show the 
subdiffusion of energy in Sec. \ref{ssec:subdiff}, and thereafter in Sec. \ref{ssec:nilm} proceed towards extracting a 
consistent explanation of the thermal transport properties using the concept of intrinsic localized modes of excitation 
in discrete nonlinear systems. We extend our analysis to the $\delta > 0$ regime for the LR-XY model in Sec. \ref{ssec:delta} 
and conclude with a discussion in Sec. \ref{conclusions}.


\section{The model}
\label{sec:model}
The Hamiltonian mean-field model consists of $N$ particles (or classical planar spins) on a one-dimensional lattice that interact with each other
following the energy functional
\begin{equation}
\mathcal{H} = \sum_{i = 1}^N \frac{p_i^2}{2} + \frac{1}{2 N} \sum_{i,j = 1}^N 1 - \cos(\theta_j - \theta_i),
\label{H}
\end{equation}
where $p_i$ and $\theta_i$ are the conjugate momentum and position (on a circle) of the $i$th particle. As before, the $1/N$ factor in the
potential energy makes it (and consequently the total energy) extensive, which is the so-called Kac prescription. For the time being we impose
periodic boundary conditions. The equation of motion for each particle can be conveniently written in terms of the magnetization components
$M_x = \frac 1N \sum_{i=1}^N \cos \theta_i$ and $M_y = \frac 1N \sum_{i=1}^N \sin \theta_i$ as 
\begin{equation}
\dot{p_i} = F_i = M_y \cos \theta_i - M_x \sin \theta_i,
\label{EoM1}
\end{equation}
where $F_i = - \frac{\partial V_i}{\partial\theta_i}$ is the force and $V_i = \frac 1N \sum_j [1- \cos(\theta_j - \theta_i)]$
is the potential experienced by the $i$-th particle. The share of the total energy that the $i$-th particle gets is
\begin{equation}
E_i = \frac{p_i^2}{2} + \frac{1}{2 N} \sum_{j = 1}^N 1 - \cos(\theta_j - \theta_i).
\label{E_i}
\end{equation}
We investigate the properties of this model by numerically integrating the equations of motion using the symplectic velocity Verlet integrator
\cite{vel-Verlet} with a small time-step $\Delta t = 0.01$. The initial conditions are always assigned randomly from a uniform distribution
for the angle variables and from a Gaussian distribution with unit variance for the momenta, both centered around zero. Our results are presented
in the next section.


\section{Simulation results}
\label{sec:results}

\subsection{Nature of energy diffusion}
\label{ssec:subdiff}
First we verify the nature of energy diffusion in the HMF model. Following the equilibrium fluctuation-correlation method, we start from an
equilibrated microcanonical system under periodic boundary conditions and monitor how the excess energy at site $i$ at time $t_0$ propagates
to site $j$ at a later time $t_0 + t$. This is achieved by studying the spatio-temporal correlation of the energy fluctuations which for a
microcanonical system (see Refs. \cite{FC1,FC2} for details) is
\begin{equation}
\rho_{_E}(r,t) = \frac{\langle \Delta E_j(t+t_0) \Delta E_i(t_0) \rangle}{\langle \Delta E_i(t_0) \Delta E_i(t_0) \rangle} + \frac 1 {N_b - 1}.
\label{rho_E}
\end{equation}
As is usually done, we have coarse grained the lattice into $N_b = N/b$ bins with $b$ number of particles in each bin and $r = (i-j)b$. Here
$\Delta E_k = E_k - \langle E_k \rangle$ is the excess energy of the $k$th bin ($1 \le k \le N_b$); $E_k$ ($\langle E_k \rangle$) is the
instantaneous (average) energy of all the $b$ particles in the $k$th bin calculated using Eq. (\ref{E_i}). For our numerical results we set $b = 4$.
The nature of the energy diffusion process is ascertained from the mean square deviation (MSD) of the function $\rho_{_E}(r,t)$ denoted as
\begin{equation}
\sigma_{_E}^2 (t) = \sum_{r = - N/2}^{N/2} r^2 \rho_{_E}(r, t).
\label{msd}
\end{equation}
For a subdiffusive (superdiffusive) process we should have $\sigma_{_E}^2 (t) \sim t^{\beta}$ where $\beta < 1$ ($\beta > 1$) at large times 
$t$; $\beta = 1$ corresponds to normal diffusion. After initialization, the momenta are rescaled so that the system has the desired energy
density $u = \langle \mathcal{H}\rangle/N$ and zero net momentum $\sum_i p_i = 0$. We evolve this system numerically for a long time $t = 10^6$
($t$ is measured in units of $\Delta t$ here) until it reaches stationarity. The equilibrated system should have a statistically constant energy 
profile with $E_i = u$. Once this is attained, we start computing the excess energy correlation function $\rho_{_E}(r,t)$, and repeat this over
several ($\sim 10^6$) initial values of $t_0$ until well-averaged data are obtained for $\sigma_{_E}^2 (t)$.
\begin{figure}[htb]
\centerline
{\includegraphics[width=5.25cm,angle=-90]{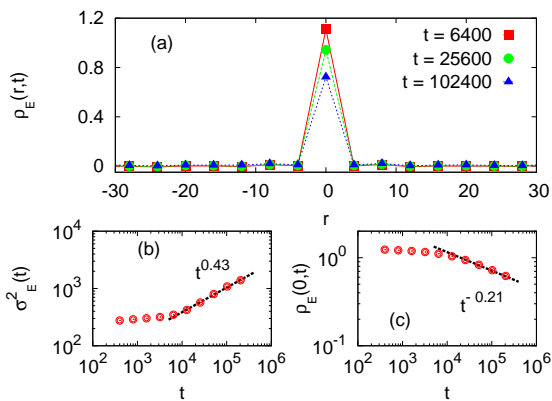}}
\caption{Subdiffusion of energy in the HMF model for $N = 200$ and $u = 1.0$: (a) the correlation function $\rho_{_E}(r,t)$ at different times
$t$ (only a small window around $r = 0$ is shown here), (b) the MSD grows as $\sigma^2_{_E} (t) \sim t^{0.43}$ and (c) the height decreases with
time as $\rho_{_E}(0,t) \sim t^{-0.21}$ at large times indicating energy subdiffusion.}
\label{fig:subdiff}
\end{figure}
We choose a HMF system of size $N = 200$ which is the largest $N$ that is studied in Ref. \cite{LRXY} with an energy density $u = 1.0$.
The correlation function $\rho_{_E}(r,t)$ at different times is shown in Fig. \ref{fig:subdiff} (a). It can be seen that the spread of
fluctuations is extremely slow, $\rho_{_E}(r,t) \approx 0$ for $r \neq 0$ even for very large $t > 10^5$. The MSD $\sigma_{_E}^2 (t)$
is displayed in Fig. \ref{fig:subdiff}(b); at large times, $\sigma^2_{_E} (t) \sim t^{0.43}$. The height $\rho_{_E}(0,t)$ of the correlation
function decreases with time as $\rho_{_E}(0,t) \sim t^{-0.21}$, thus much slower than what one would expect for normal diffusion,
i.e., $\rho_{_E}(0,t) \sim t^{-1/2}$. 
Thus the diffusion process in the HMF model is indeed subdiffusion with $\beta < 1$ and the thermally driven system should therefore be an
insulator in the thermodynamic limit, as was predicted for a subdiffusive system \cite{Subdiff1} and numerically demonstrated in Ref. \cite{LRXY}.
This is intriguing, as mentioned earlier, that LR interactions enhance thermal transport in the LR-FPU model whereas suppress  the same in the
LR-XY model.

\begin{figure}[htb]
\centerline
{\includegraphics[width=5.25cm,angle=-90]{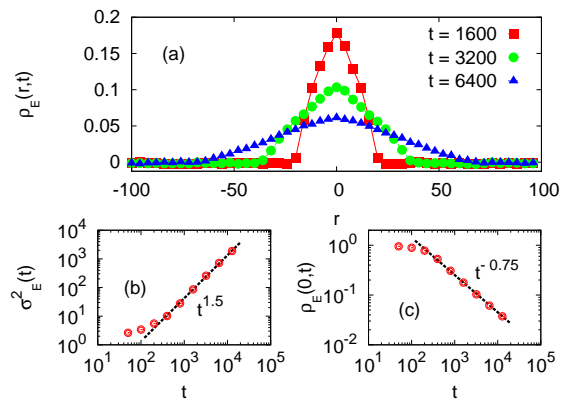}}
\caption{Superdiffusion of energy in the GHMF model for $N = 200$ and $u = 1.0$: (a) the correlation function $\rho_{_E}(r,t)$ at different 
times $t$, (b) the MSD grows as $\sigma^2_{_E} (t) \sim t^{1.5}$ and (c) the height decreases with time as $\rho_{_E}(0,t) \sim t^{-0.75}$ 
at large times indicating energy superdiffusion.}
\label{fig:supdiff}
\end{figure}

Next, note that the Hamiltonian for the LR-FPU model also has NN interactions [second term in Eq. (\ref{H1})], besides the algebraically
decaying LR term. In the LR-XY Hamiltonian Eq. (\ref{H2}) a similar NN interaction term is not present. In order to study the effect of
additional NN interactions on the nature of energy diffusion, we modify the HMF Hamiltonian as
\begin{equation}
\mathcal{H}_G = \mathcal{H} + \sum_{i = 1}^N 1-\cos(\theta_{i+1} - \theta_i),
\label{H_G}
\end{equation}
where $\mathcal{H}$ is given by Eq. (\ref{H}). The HMF model with additional NN interactions has been studied earlier \cite{GHMF}
and is referred to as the generalized HMF (GHMF) model in the following. The equation of motion now has the form 
\begin{equation}
\dot{p_i} = M_y \cos \theta_i - M_x \sin \theta_i + \sin(\theta_{i+1} - \theta_i) + \sin(\theta_{i-1} - \theta_i)
\label{EoM2}
\end{equation}
To verify if the presence of the NN term leads to a superdiffusive behavior similar to the LR-FPU model we compute, as before, the
correlation $\rho_{_E}(x,t)$ for different times, which is shown in Fig. \ref{fig:supdiff}(a) for the same parameters as in 
Fig. \ref{fig:subdiff}. As can be immediately appreciated, the spreading of the correlation function is much faster as compared 
to the HMF model. At large times the MSD shows $\sigma^2_{_E} (t) \sim t^{1.5}$ and the height decreases with time as 
$\rho_{_E}(0,t)\sim t^{-0.75}$, as is shown in Figs. \ref{fig:supdiff}(b) and \ref{fig:supdiff}(c), respectively. Thus a
clear superdiffusion is observed at large times with $\beta > 1$ in this case. 
This demonstrates that adding a NN interaction term to the HMF model {\it speeds up} the diffusion process from subdiffusion to superdiffusion.
This seems to be counter-intuitive since adding nonlinear interactions usually leads to more scattering of the heat carriers, thus making the
transport process slower. As a simple example, adding quartic interactions to the Harmonic oscillator model slows down the diffusion process
from ballistic ($\beta = 2$) to superdiffusion ($1< \beta < 2$) as in the FPU model \cite{HT_Dhar}. 

\begin{figure}[htb]
\centerline
{\includegraphics[width=4.85cm,angle=-90]{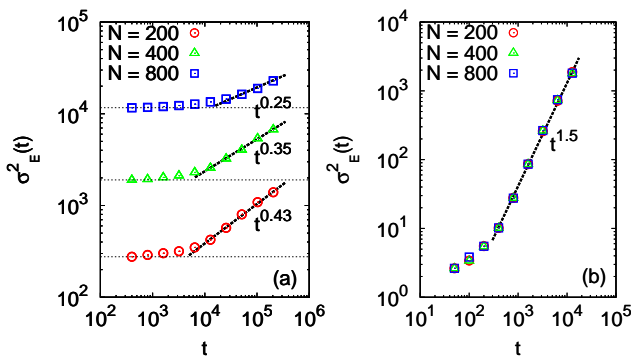}}
\caption{The MSD $\sigma^2_{_E} (t)$ as a function of time $t$ with $N = 200, 400$, and $800$ for (a) the HMF model and (b) the 
GHMF model. The energy density is fixed at $u = 1.0$.} 
\label{fig:msd_N}
\end{figure}

We conclude this section by verifying that the nature of energy diffusion for the two models does not change, at least qualitatively, as
the system size $N$ is increased. In Fig. \ref{fig:msd_N}(a) the MSD $\sigma^2_{_E}(t)$ of the HMF model is shown for three different system 
sizes, $N = 200, 400$, and $800$. Up to very large times $t > 10^5$ that we have computed, the $\sigma^2_{_E}(t)$ for the larger $N$ is found
to have a slower increase with $t$ as compared to that of a smaller $N$ value: For $N = 200, 400$, and $800$ we obtain $\beta \approx 0.43, 
0.35$, and $0.25$, respectively. Moreover, the point of inflection for the three curves moves rightwards to larger $t$ values as $N$ is increased.
Thus a slowdown of energy diffusion happens with increasing $N$, which indicates that, in the thermodynamic limit, the energy diffusion will
remain subdiffusive for the HMF model. 
In fact, such a slowing down of the dynamics with increasing system size is well known for this model. It has been observed that relaxation of the
HMF model towards the Boltzmann-Gibbs equilibrium becomes exceedingly delayed as $N$ is increased.  This is because the system stays trapped in
the so-called {\it quasi-stationary} states and the relaxation time-scale generally diverges faster than $N$ \cite{LR_Ruffo, LR_Mukamel}.

On the other hand, for the GHMF model, the exponent $\beta$ seems to have negligible finite-size effects, as depicted in Fig. \ref{fig:msd_N}(b).
The MSD $\sigma^2_{_E}(t)$ for the three system sizes, $N = 200, 400$, and $800$, shows identical divergence with an exponent $\beta \approx 1.5$
at large times. Hence we speculate that the energy diffusion in the GHMF model will remain superdiffusive, even in the thermodynamic limit,
akin to the LR-FPU model \cite{LRFPU}. In the next section we provide more numerical evidence to justify these results further.


\subsection{Nonlinear intrinsic localized modes}
\label{ssec:nilm}
In order to explain the energy subdiffusion in the HMF model and why additional NN interactions give rise to superdiffusion, we now
study certain energy relaxation processes in the HMF (and also the GHMF) model. This is done by performing {\it boundary cooling} experiments
in which one studies how a system relaxes from a high-temperature equilibrium state to a low-temperature equilibrium state in the presence
of {\it absorbing boundaries} \cite{DB1,DB2,DBLivi}. In the following, we describe the cooling protocol that has been adopted here.

We begin by immersing our desired model (the HMF or the GHMF with free boundary conditions) in a heat bath at a fixed temperature $T_b$
which is set to a relatively high value. The stochastic heat bath is modeled by the classical Langevin model \cite{HT_Dhar}. In other words,
to {\it each} particle in the lattice we apply Gaussian white noise and friction (dissipation) at a temperature $T = T_b$. The equations of
motion become
\begin{equation}
\dot{p_i} = F_i - \gamma p_i + \sqrt{2 \gamma k_B T_b} ~\eta_i,
\end{equation}
where $\gamma$ is the friction coefficient and $\eta_i$ is a delta correlated Gaussian white noise with zero mean (the friction coefficient
$\gamma$ and the Boltzmann constant $k_B$ are set to unity in all our results).
Once the system gets equilibrated under this process at temperature $T = T_b$, we remove the system from the heat bath and thereafter apply
friction {\it only} to the two ends of the system. Equivalently, this can be understood as attaching two Langevin heat baths at temperature
$T = 0$ to the end particles at sites $i = 1$ and $i = N$. Explicitly written, the equations of motion now become
\begin{equation}
\dot{p_i} = F_i - \gamma p_i (\delta_{i1} + \delta_{iN}),
\label{cool}
\end{equation}
where $\delta_{ij}$ is the Kronecker delta function. The system will start to cool off gradually from the equilibrium state at temperature 
$T = T_b$ to the equilibrium (ground) state at $T = 0$ by dissipating heat from the two ends. We allow this boundary cooling to happen for
a very long time to allow the system to relax to the ground state.

At this point one might expect that the entire system will relax to the ground state, within a reasonably short time-scale, to have a 
homogeneous temperature and energy density, consistent with zero temperature. This, however, is not generally true in discrete nonlinear 
lattices because of the spontaneous thermally generated intrinsic localized modes of excitations often referred to as discrete breathers
(DBs) (for recent reviews, see Refs. \cite{FlachRev1998,AubryRev2006, FlachRev2008}). These are dynamical structures which have been seen in many
nonlinear discrete systems, classical and quantum \cite{ClQnt}, in equilibrium and nonequilibrium, and in any dimension under generic
conditions, although their specifics, such as energy thresholds, stability, mobility, etc., may depend on the details of the system in which
they appear. These discrete breathers trap a significant fraction of the system's energy for a long time but are eventually destroyed in
finite systems, thus releasing their trapped energy. In fact, one of the possible explanations for observing normal heat conductivity in the
inertial XY model \cite{Rotor1} instead of superdiffusive transport, despite being momentum conserving, is attributed to these localized
``rotobreathers'' preventing the free propagation of phonons regardless of their wavelength or frequency \cite{Rotor2, Rotor3}. Thus these
cooling experiments are an efficient method of detecting the presence of localized excitation modes and studying their properties.

Here, in the following, we will apply the idea of these intrinsic localized modes to systems with LR interactions. In LR systems, with
algebraically decaying interaction strengths, DBs have been studied as mathematical objects \cite{DBLR0}, but their role in influencing
the macroscopic properties, such as relaxation and energy transport, has yet to be investigated. We show that energy localization due
to the emergence of these nonlinear excitations can consistently explain all the results discussed above.
\begin{figure}[htb]
\centerline
{\includegraphics[width=5.cm,angle=-90]{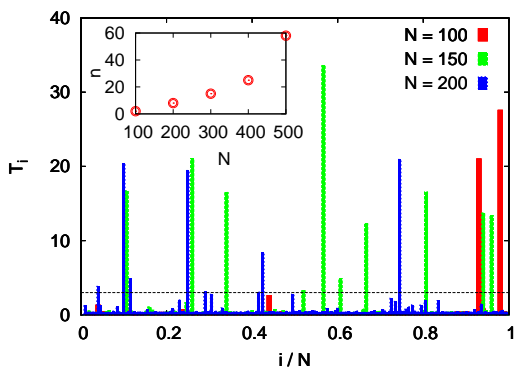}}
\caption{The temperature profile $T_i$ of the HMF model for different system sizes $N = 100, 150$, and $200$. The $x-$axis represents 
the particle indices $i$ and has been rescaled by the system size $N$. The dashed horizontal line indicates the initial temperature
$T_b = 3.0$. Inset: The number of localized excitations $n$ having $T_i > T_b = 3.0$ increases as the system size $N$ is increased.}
\label{fig:db}
\end{figure}

Starting from an initial high temperature $T_b = 3.0$, we first thermalize the HMF chain. For computational convenience, we intentionally
choose a high initial temperature since these intrinsic nonlinear modes are thermally generated excitations and therefore more pronounced 
at elevated temperatures \cite{DB1, DB2}. After thermalization, we employ the boundary cooling which eliminates most of the delocalized 
excitations, such as phonons and solitary waves, leaving behind only the localized excitation modes in the system. At this point, we start
computing the time-averaged temperature $T_i = \langle p_i^2 \rangle$ of each particle (or, equivalently, the particle's energy $E_i$) of
the system.

Our result from the boundary cooling experiment is depicted in Fig. \ref{fig:db}.
We find that, even after allowing an appreciably large time ($t > 5 \times 10^7$) for the system to attain the ground state, there are certain 
localized regions in the lattice that have temperatures even higher than the initial temperature $T_b = 3.0$ at which we had equilibrated our
system in the beginning. These ``hot spots'' trap a large amount of energy and are extremely long-lived, thus indicating that energy diffusion 
in the lattice is extremely slow or otherwise these localizations would have disappeared because of diffusion of energy from the high-energy to
low-energy points in the lattice. This also makes it apparent why the HMF model in the thermodynamic limit behaves as an insulator under thermal
bias and exhibits energy subdiffusion. As the system size increases, a large number of these localizations appear and this blocks the passage
of energy and renders the lattice non-conducting. Under identical simulation conditions, the number $n$ of sites with temperature $T_i > 3.0$
(indicated by the dashed line in Fig. \ref{fig:db}) increases with the system size $N$, as can be seen in the inset of Fig. \ref{fig:db}. Thus,
in the thermodynamic limit, the entire lattice essentially splits up into a large number of localized points with negligible interactions among
them, which gets reflected also in the result that the correlation $\rho_{_E}(r,t) \approx 0$ for $r \neq 0$ that we obtained in the previous
section in Fig. \ref{fig:subdiff}(a). This also helps us to understand why in Fig. \ref{fig:msd_N}(a) the MSD $\sigma^2_{_E}(t)$ shows a
slower increase with time $t$ (smaller $\beta$ exponent) when $N$ is increased.

\begin{figure}[htb]
\centerline
{\includegraphics[width=3.75cm,angle=-90]{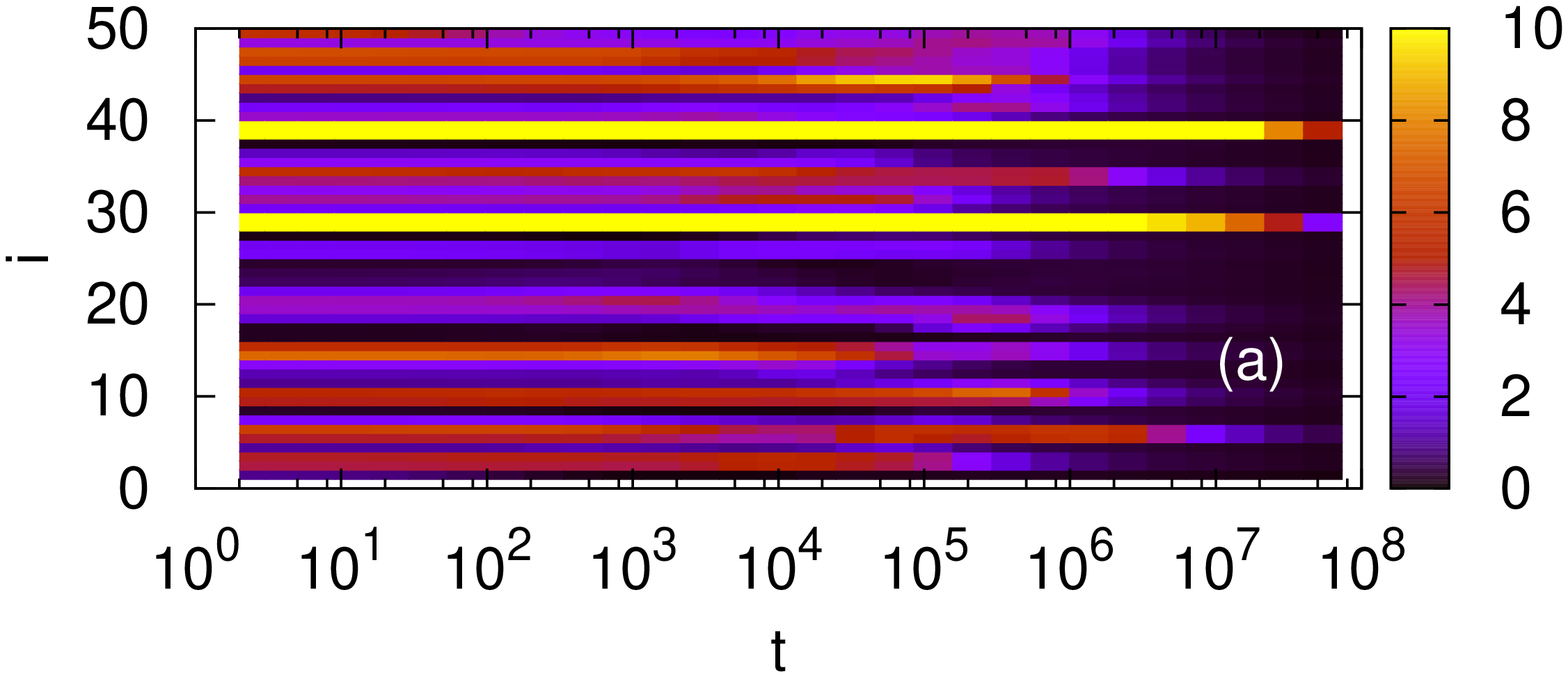}}\vskip-0.35cm
{\includegraphics[width=3.75cm,angle=-90]{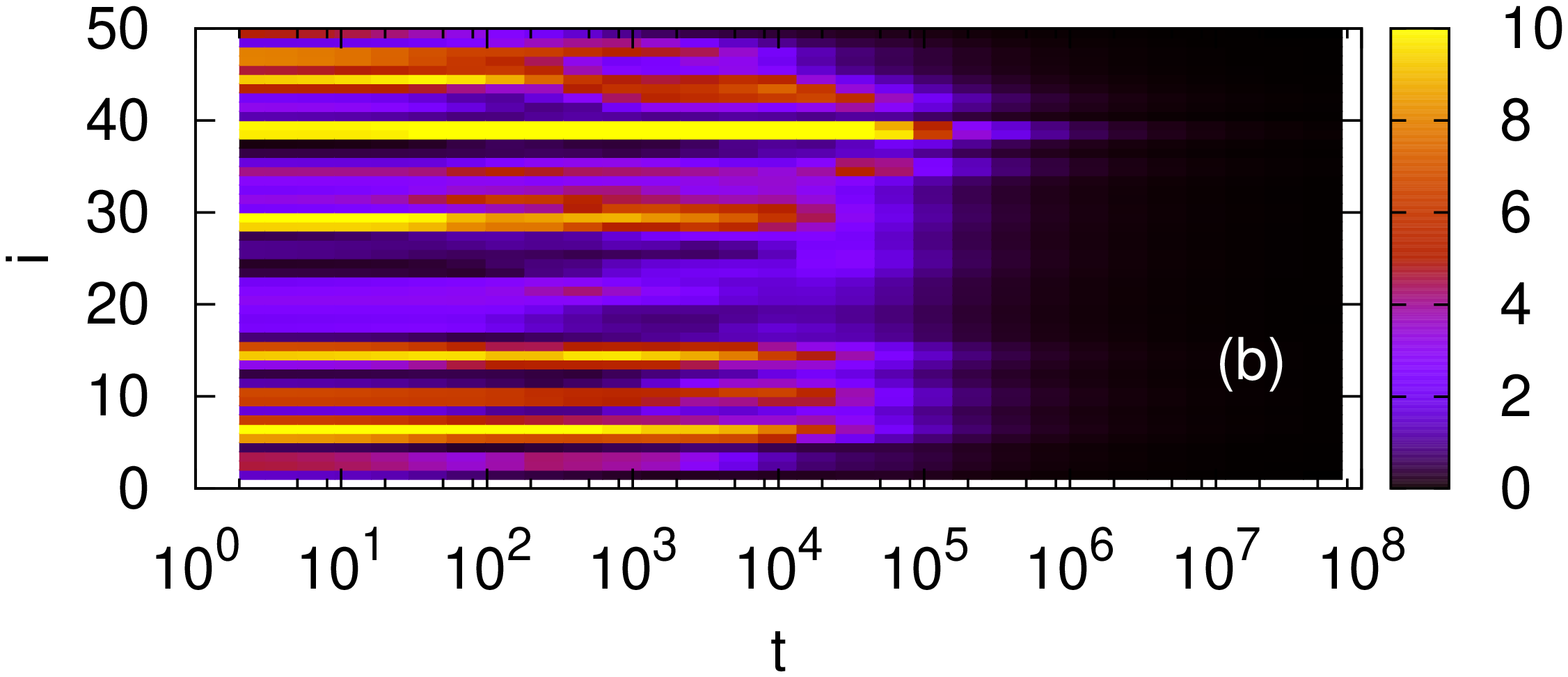}}\vskip-0.35cm
{\includegraphics[width=3.75cm,angle=-90]{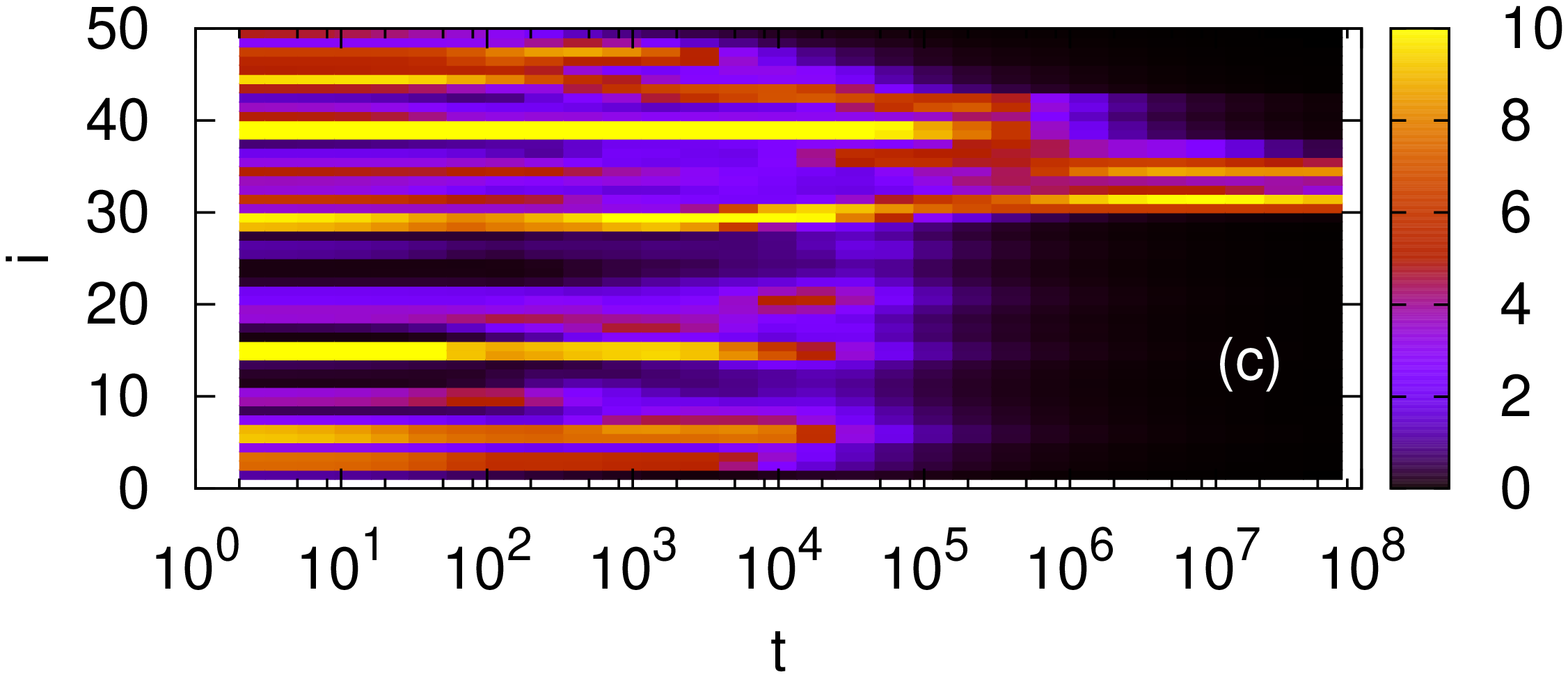}}
\caption{Variation of the  particles' temperature with time $t$ for (a) the HMF model, (b) the GHMF model, and (c) the inertial XY (coupled rotor)
model, all for $N = 50$. The $y-$axis represents the particle indices on the lattice denoted by $i$, and the $x-$axis represents time $t$. The colors
indicate the average temperature $T_i$ of each particle ranging from $T_i = 0$ to $T_i = 10$ as shown in the scale beside each figure.}
\label{fig:cool}
\end{figure}

In Fig. \ref{fig:cool}(a), we show how the local temperature evolves with time in a HMF lattice of size $N = 50$ after the boundary cooling is 
started and the system is allowed to relax to the ground state. We observe that the localizations persist beyond very large times $t = 10^7$ 
even in a system of such a small size.

When we perform the same cooling protocol on the GHMF model, we find that the localization effect vanishes at large times. This can be
seen by comparing Fig. \ref{fig:cool}(a) and Fig. \ref{fig:cool}(b) - the latter appears to be statistically homogeneous at a much shorter
time-scales when compared to the former. This shows that, even if energy gets localized initially at some points in the  lattice, the presence
of the NN interactions makes them unstable. The trapped energy is released and efficiently diffused uniformly in the entire lattice by the
LR interactions. This fast annihilation of the energy localizations is negligible in the HMF model because the two-body LR interactions
are extremely weak (by a factor of $N^{-1}$ due to the Kac prescription) as compared to the NN interaction term in the GHMF model. One can
clearly see that the diffusion of energy from one site to its neighbors is negligible in the case of the HMF model when compared to the
GHMF model.

Now, we compare the results of the HMF and GHMF with the case of the nearest-neighbor XY (rotor) model, which is shown in Fig. \ref{fig:cool}(c). 
We perform boundary cooling on the XY model with the same set of parameters and find that energy localization in this case is weaker (having
more sites with $T_i \approx 0$ at large times) as compared to that in the HMF model, but stronger than the GHMF model.

From these results, we can intuitively understand that LR interactions have two competing effects. One is, as we expect, transporting energies
over long distances and thus making energy transport (and equilibration) extremely fast. On the other hand, in the presence of LR interactions
energy localization occur abundantly and are extremely long lived. These localized excitation modes impede energy transport in {\it pure}
LR models such as the HMF and exhibit energy subdiffusion. The addition of NN interactions introduces strong local perturbations that
cause these localizations to annihilate rapidly. Thus in the presence of both LR and NN interactions, the energy relaxation is the fastest and
transport is superdiffusive as in the GHMF model. This justifies why one obtains $\beta_{HMF} < \beta_{XY} = 1 < \beta_{GHMF}$, as indicated
by our energy diffusion results. These results also make it very clear why introducing LR interactions in the FPU model enhances its
thermal conduction whereas it turns the XY model into a thermal insulator.

\subsection{The $\delta > 0$ regime of the LR-XY model}
\label{ssec:delta}
Until now we have focused mostly on the HMF model which is the $\delta = 0$ case of the LR-XY model Eq. (\ref{H2}). Let us now extend the
idea of localized modes to finite values of $\delta > 0$. In the energy transport study \cite{LRXY} it was found that the conductivity
$\kappa$ monotonically increases with the increase of $\delta$ from zero (Fig. 2 in Ref. \cite{LRXY}). We speculate that this increase in
conductivity with $\delta$ should get reflected also in the emergence of the localized nonlinear excitation modes. This is what we study
in the following.

\begin{figure}[htb]
\centerline
{\includegraphics[width=5.cm,angle=-90]{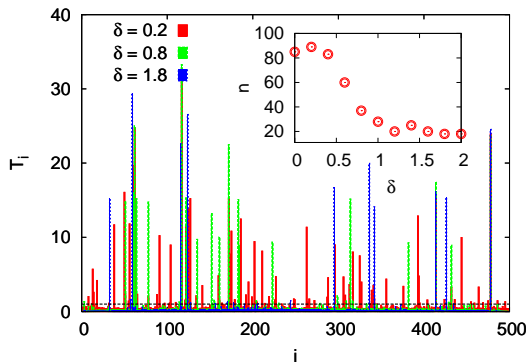}}
\caption{Temperature profile for the LR-XY model for different values of $\delta$ with $N = 500$. Inset: The number $n$ is
estimated as a function of $\delta$ with the temperature threshold set at $T_{th} = 1.0$ represented by the horizontal dashed
line in the main figure.}
\label{fig:DBa}
\end{figure}

As before, starting from random initial conditions, we thermalize a LR-XY system with $N = 500$ particles, by immersing it in a heat bath
at temperature $T_b = 3.0$. Thereafter we allow it to cool off for a large enough time $t = 10^6$ using boundary dissipation and compute
the local temperature $T_i$ of the system, from which we will have an estimate of the relative number of localized modes for different
values of $\delta$. Since the system at large times should ideally have zero temperature at all sites, we set a threshold at a nonzero
$T_{th} > 0$ and identify any site with $T_i > T_{th}$ as a breather site. The localizations appear for $\delta > 0$, as is shown in
Fig. \ref{fig:DBa}, but their number $n$ decreases with increasing $\delta$. Under identical simulation conditions, the variation of
$n$ as a function of $\delta$ is shown in the inset of Fig. \ref{fig:DBa} with the temperature threshold set at $T_{th} = 1.0$.
We find that in the LR regime ($0 \le \delta < 1$) the number of localized excitations $n$ falls off steadily as the parameter
$\delta$ approaches unity, as shown in the inset of Fig. \ref{fig:DBa}. After crossing the LR regime at $\delta = 1$, the number
$n$ seems to be approximately constant. This decrease in $n$ increases the speed of diffusion as $\delta$ is increased and one
obtains a normal thermal conductor for $\delta > 1$ from a thermal insulator below $\delta = 1$, as was observed in Ref. \cite{LRXY}.
Note that the absolute number of localized excitations $n$ changes as one varies the thermalization time, cooling time, the threshold
$T_{th}$, initial conditions, size $N$, parameters of the system and bath, etc., which is not surprising, but the qualitative feature
(the decrease of $n$ with increase in $\delta$ from zero) of the result in Fig. \ref{fig:DBa} remains true nonetheless.


\section{Discussions}
\label{conclusions}
To summarize, we have investigated here the energy transport features of the paradigmatic HMF model and other related LR interacting
models. The energy diffusion process in the HMF model is shown to be subdiffusive, whereas an additional NN interaction term
leads to superdiffusion, thus re-validating the results obtained in the LR-XY model and the LR-FPU model. We have verified that this
result remains true in the LR-FPU model at $\delta = 0$ as well: If the NN interactions in Eq. (\ref{H1}) are switched off, one
obtains an energy subdiffusion akin to the HMF model, thus reinforcing our understanding of the role played by the NN interactions.

Using boundary cooling experiments, we studied the role played by the nonlinear localized modes in influencing energy diffusion
and transport in the presence of both NN and LR interactions. We have shown that these modes in the HMF model persist for large times
and proliferate as the system size is increased, which ultimately make the model a thermal insulator in the thermodynamic limit. In
the presence of additional NN interactions these localized modes annihilate comparatively faster and thus lead to superdiffusive
transport as in the GHMF.
As the LR parameter $\delta$ is increased from zero in the LR-XY model, the number of these localized excitations decreases and
this changes the thermal transport from insulating to conducting (obeying Fourier's law) for large $\delta > 1$. 
Note that the boundary cooling experiment does not {\it generate} the localized excitations but rather {\it reveals} them by
removing the nonlocalized excitations from the system.
We believe that a more detailed analysis, taking into account also the mobility of these modes, could be more instructive
in understanding thermal transport properties.

Intrinsic localized modes have been observed experimentally in a variety of systems, such as the Josephson junctions, nonlinear
optical waveguides, and single crystals \cite{FlachRev2008}. Recently, the HMF Hamiltonian has also been employed to model an
experimental system with atoms trapped in an optical cavity \cite{Ruffo2017Rev}. It would be fascinating to observe these
localized modes experimentally in a real physical system described by the HMF Hamiltonian.

\vskip1cm
{\bf Acknowledgments:} The author thanks Y. Levin and R. Pakter for reading the manuscript and an anonymous referee for 
helpful suggestions. This work is financially supported by CNPq (Brazil).


\end{document}